\documentclass[review]{elsarticle}
\usepackage{siunitx}
\usepackage{xcolor}
\usepackage{upgreek}
\usepackage{lineno,hyperref}
\modulolinenumbers[5]

\journal{Journal of \LaTeX\ Templates}
\newcommand{\ca}[1]{{#1}}

\begin{document}

\begin{frontmatter}

\title{\ca{Very weak carbonaceous asteroid simulants I: mechanical properties and response to hypervelocity impacts}}

\author[oca,kent]{Chrysa Avdellidou\corref{mycorrespondingauthor}}
\cortext[mycorrespondingauthor]{Corresponding author}

\author[3sr]{Alice DiDonna}
\author[class]{Cody Schultz}
\author[3sr]{Barth\'el\'emy Harthong}
\author[kent]{Mark C. Price}
\author[3sr]{Robert Peyroux}
\author[class]{Daniel Britt}
\author[kent]{Mike Cole}
\author[oca]{and Marco Delbo'}

\address[oca]{Universit\'e C\^ote d'Azur, Observatoire de la C\^ote d'Azur, CNRS, Laboratoire Lagrange, Blvd de l'Observatoire, CS 34229, 06304 Nice Cedex 4, France}
\address[kent]{Centre for Astrophysics and Planetary Science, School of Physical Sciences, University of Kent, Canterbury, CT2 7NH, UK}
\address[3sr]{Univ. Grenoble Alpes, CNRS, Grenoble INP, 3SR, F-38000 Grenoble, France}
\address[class]{Department of Physics, University of Central Florida, Orlando, FL, USA}

\begin{abstract}
The two on-going sample return space missions, Hayabusa2 and OSIRIS-REx are going to return to Earth asteroid regolith from the carbonaceous near-Earth asteroids Ryugu and Bennu. The two main processes that lead to regolith production are the micrometeorite bombardment and the thermal cracking. 
Here we report the production of a weak simulant material, analogue to carbonaceous meteorites with a CM-like composition, following the preliminary compositional results for Bennu and Ryugu. This asteroid simulant has compressive and \ca{flexural} strength 1.8 $\pm$ 0.17 and 0.7 $\pm$ 0.07~MPa, respectively. The thermal conductivity (in air) of the simulant at room temperature is between 0.43 and 0.47~W~m$^{-1}$~K$^{-1}$. In order to distinguish the type of regolith that is produced by each of these processes, we present and discuss the results of the experimental campaign focused on laboratory hypervelocity impacts, using the 2-stage light-gas gun of the University of Kent, that mimic the micrometeorite bombardment.
We find that this process produces both monomineralic and multimineralic fragments, resulting in a difficulty to distinguish the two processes, \ca{at least on these weak materials}.  

\end{abstract}


\end{frontmatter}


\section{Introduction}
\label{intro}
``Extremely weak, extremely rare on Earth, but probably very common in space". In the light of the results of the two on-going sample return space missions, JAXA's Hayabusa2 \citep{2019Sci...364..272K} and NASA's OSIRIS-REx \citep{2019Natur.568...55L}, the above statement appears to be an appropriate definition for the type of materials that constitute most of the boulders on the surface of the near-Earth asteroids (162173) Ryugu \citep{grott2019,2019Sci...364..268W,sugita2019} and (101955) Bennu \citep{2019NatAs...3..341D,walsh2019}. 

In the case of Ryugu, information about the weak strength of the boulders was essentially derived from measurements, obtained by the MARA radiometer \citep{grott2017} of the thermal conductivity $\kappa$ of the solid boulder material on which MASCOT, the Hayabusa2 lander \citep{ho2017}, came to rest. 
In addition, observations of the entire surface of Ryugu in the thermal infrared indicate that the global value of Ryugu thermal inertia ($\Gamma$), a parameter that measures the resistance of a surface to temperature changes \cite[][and references therein]{Delbo2015aste.book..107D}, lies between 200 and 500~J~m$^{-2}$~K$^{-1}$~s$^{-1/2}$\citep{sugita2019} and is thus very similar to the value measured for the boulder observed by MARA \citep[$\Gamma$ = 282~$^{+93}_{-35}$J~m$^{-2}$~K$^{-1}$~s$^{-1/2}$;][]{grott2019}. 
Since the surface is essentially covered by boulders \citep{sugita2019}, this gives an indication that the typical thermal inertia of Ryugu's rocks is similar to that measured by \cite{grott2019}. $\Gamma$ is defined as the square root of the product of thermal conductivity $\kappa$, density, and heat capacity of a material, and thus knowledge of the value of $\Gamma$ allows the estimation of $\kappa$. Under reasonable assumptions of the material density and heat capacity, the value of $\kappa$ for Ryugu's boulder was found to be lower than expected, namely several factors lower than those measured in the laboratory for meteorites \citep{opeil2010, opeil2012}. Specifically, $\kappa$ was calculated to be between 0.06 and 0.16~W~m$^{-1}$~K$^{-1}$ (at 230~K), whilst meteorites range between 0.45--5.5~W~m$^{-1}$~K$^{-1}$ at 200~K \cite[][and references therein]{opeil2012}. The lower thermal conductivity of Ryugu's boulders compared to those of meteorites can be explained as an effect of the higher porosity of Ryugu's boulder material compared to the meteorites \citep{grott2019}. Indeed, it should be noted that our sample of asteroid meteoritic material is likely biased, as we could expect the weakest meteoroids to be unable of surviving passage through Earth's atmosphere. 

Initial ground-based spectroscopic observations of Ryugu indicated a CM or CI-type composition \citep{moskovitz2013,perna2017,lecorre2018}. Later remote spectral observations of Ryugu from Hayabusa2 confirmed the initial ground-based results \citep{2019Sci...364..272K}, however it was not possible to derive a definite meteoritic analogue. First of all there are no measured meteorite spectra that match perfectly the measured visible and near-infrared spectra of Ryugu in the wavelength range between 0.45 and 3~$\upmu$m. The closest match is produced by thermally-metamorphosed CI chondrites and shocked CM chondrites \citep{2019Sci...364..272K}.  Additionally, close inspection of Ryugu's boulder images revealed mm-size inclusions \citep{grott2019}. CM and CM2 chondrites, according to our current meteoritic collection have smaller inclusions \cite[e.g. chondrules of 0.27--0.3~mm,][]{burbine2016}, making a link between Ryugu's boulder and said meteorites possibly problematic. On the other hand, CI meteorites have less than 1\% calcium aluminium-rich inclusions and chondrules \citep{burbine2016}.

Using Hertzian heat conduction theory, MASCOT team \citep{grott2019} used the value of $\kappa$ and the estimated boulder porosity to evaluate the mechanical tensile strength of the material, which was calculated to be \ca{between 0.20 and 0.28~MPa}. This is weaker than typical meteoritic material \cite[and references therein]{ostrowski2019}, including the most fragile CM and CI carbonaceous meteorites: for instance, the compressive strength of the CM2 Murchison and Sutter's Mill meteorites are around 50~MPa \citep{miura2008} and 85~MPa \citep{jenniskens2012} respectively. Following common assumption that compressive strength is approximately 10 times the tensile strength, one can estimate that the tensile strength of the aforementioned meteorites, is at best, a few MPa. Observations of the ungrouped C2 Tagish Lake fireball, \citep{brown2002} estimated that the pre-atmospheric compressive strength was about 0.25~MPa, while its material compressive strength is 0.7~MPa, the latter being more appropriate for the recovered meteorites. Later laboratory experiments on samples of Tagish Lake give a tensile strength of 0.8 $\pm$ 0.3~MPa \citep{tsuchiyama2009}. The combination of these two independent measurements, makes Tagish Lake a peculiar case, as its compressive and tensile strength have very similar values. Furthermore, the CI Ivuna and Orgueil have tensile strengths of 0.7 $\pm$ 0.2 and 2.8 $\pm$ 1.9~MPa respectively, as measured by \cite{tsuchiyama2009}, while the CM2 Murray have a value of 8.8 $\pm$ 4.8~MPa, according to measurements from the same authors. 

OSIRIS-REx provided similar, albeit less constraining, observations compared to the ones reported by \cite{grott2019}, but still favouring the interpretation that places Bennu's materials within the weakest known meteorites. In particular, thermal infrared data allowed \cite{2019NatAs...3..341D} to determine the thermal inertia of Bennu to be 350 $\pm$ 20~J~m$^{-2}$~s$^{-0.5}$~K$^{-1}$, corresponding to thermal conductivities very similar to those derived by \cite{grott2019} for Ryugu. However, these $\kappa$ values for Bennu, were determined from the thermal signal of the entire visible and illuminated surface of Bennu and not just a single boulder. Nevertheless, boulders constitute a large fraction of Bennu's surface \citep{2019NatAs...3..341D,walsh2019}, implying that they could have low $\kappa$, large porosity, and therefore low mechanical strength.

For Bennu, initial ground-based spectroscopic observations classified the body as a B-type asteroid with a composition more similar to the CM1 meteorites \citep{clark2011}. OSIRIS-REx observations confirmed the blue slope that was previously detected and additionally showed an absorption at around 2.74~$\upmu$m \citep{hamilton2019}, interpreted with the presence of hydrated materials. The exact position of this feature gives more information on the composition, where in this case is close to CM2.1--CM2.2 meteorites \citep{hamilton2019}. Unlike Ryugu, Bennu appears strongly hydrated and less heated, consistent with their estimated orbital evolutions \citep{Michel2010Icar..209..520M,Delbo2011ApJ...728L..42D}. 

Concerning the abundance of these materials, it has been estimated that Ryugu and Bennu are very likely originated from the re-accumulation of debris of larger parent asteroids fragmented by impacts with other asteroids in the region of the Main Belt that is closer to the Sun (2.1 $<$ a $<$ 2.5~au) and comprises asteroids with low orbital inclination i, e.g. $i<\sim7^{\circ}$ \citep{bottke2015,deleon2016,campins2010,Campins2013AJ....146...26C}. This area is populated by a large amount of asteroids with spectral and albedo properties similar to those of Ryugu and Bennu \citep{delbo2017,deleon2016,walsh2013} implying that their materials could be common in the Main Belt. On the other hand, because of their fragile nature, these materials are extremely rare on Earth as they filtered by the atmosphere: meteoroids carrying fragile materials fragment in the high atmosphere, resulting in the lack of meteorites that could represent Ryugu's and Bennu's materials.

Even though Ryugu and Bennu have landscapes covered by boulders in the meter-size range, areas with small particles in the size ranges of some cm are reported from the images returned by Hayabusa2 (during the operations approaching the surface) and OSIRIS-REx \citep{2019Natur.568...55L}. Both Hayabusa2 and OSIRIS-REx will return samples of their regoliths to Earth for detailed laboratory analysis. In particular, the OSIRIS-REx is due to return a minimum of 60~gr of regolith, offering a unique opportunity to analyse a large amount of low-albedo asteroid material \citep{lauretta2015}. This sample will allow to study the bulk properties of an asteroidal regolith. One of the central aims of the research related to these missions is to understand how the regolith of these asteroids formed, evolved, and how it is related with the interiors of these asteroids. 

As on our Moon, it was thought that the regolith on asteroids originates from the impact of micrometeorites hitting the surface with speeds of several km/s and breaking up surface rocks into smaller pieces, a process called comminution \citep{horz1975,horzcintala1997,Basilevsky2015P&SS..117..312B}. However, recent studies have shown from experimental and modelling approaches that regolith could also be produced by another physical processes, the so-called thermal cracking \citep{delbo2014,hazeli2018,molaro2012,molaro2017}. This is a mechanical stress, which produces damage and eventually causes the failure of the surface rocks, due to the cycles of heating and cooling when the asteroid surface transitions from day to night. These predictions appear validated by images obtained during the preliminary survey of the asteroid Bennu by OSIRIS-REx, which found evidence of boulders broken in place and desegregating boulders in regolith \citep{2019Natur.568...55L,walsh2019}, and by other astronomical observations of asteroids \citep{graves2019}.

Additional laboratory experiments were used to inform thermomechanical and crack propagation models, which indicated that thermal cracks propagate preferentially around different mineral phases \citep{hazeli2018}, leading to the prediction that the mode of failure of non-homogenous rocks, such as chondrite meteorites would produce monominerallic regolith grains. Indeed monominerallic grains are the major components of the regolith sample returned by the Hayabusa mission from the asteroid (25143) Itokawa \citep{nakamura2011}. 

Conversely, the break up of rocks by the much more violent (compared to slow thermal cracking) impacts of micrometeorites should produce debris with different mineralogies in the same grains. This is because the crack propagation velocity is much faster compared to that of thermal cracking. 
\ca{Pioneering experiments by \cite{durda1999} showed mineralogical size segregation from the preferential fracture along mineral grain boundaries in hypervelocity impact experiments into olivine porphyritic basalt. Similar chemical and mineralogical segregation between matrix and chondrule minerals was seen in impact disruption of ordinary chondrite and anhydrous carbonaceous (CV Allende) meteorite specimens \citep{flynn2004}. \cite{flynn2009} also performed  a single hypervelocity impact experiment on hydrous meteorite this time, specifically on Murchison, where they measured the mass distribution of the produced fragments. \cite{michikami2007} performed a series of impact experiments where they studied the effect of target porosity and strength on the impact outcome (crater formation and ejecta speeds). However, a comparison between the debris produced by different mechanisms, namely impact experiments and thermal cycling, performed on the very same type of weak natural materials such as those inferred to be present on these low-albedo carbonaceous-like asteroids, have never been performed so far.} 

Here we perform an initial study of the effects of hypervelocity impacts on these types of materials. In this work we investigate hydrated and mechanically weak \ca{asteroid boulder simulants constructed in the laboratory to have composition similar to the CM2 carbonaceous chondrites, but mechanical strength and porosity intermediate between those that have been measured for meteorites and those inferred for low-albedo and spectrally featureless asteroids, including Bennu and Ryugu}. In section 2 we describe the mineralogy and preparation of the simulants. In sections 3 and 4 we describe their mechanical and thermal characterisation. In section 5 we explain the set up to carry out the impact experiment, along with their results. In section 6 we interpret our results in comparison to the observations obtained by Hayabusa2 and OSIRIS-REx of Ryugu and Bennu.  

\section{Preparation of simulant materials}
\label{preparation}

Our specimens are based on the UCF/DSI CM Carbonaceous Chondrite Simulant (UCF/DSI-CM-2) that is part of series of asteroid regolith simulants developed in collaboration with the University of Central Florida (UCF) and Deep Space Industries (DSI) and is now produced by UCF's Exolith Lab \citep{britt2018,boivin2018,metzger2019,britt2019}. \ca{In particular, we use the CM2 regolith simulant that is a close mineralogical match to the Murchison CM2 carbonaceous chondrite meteorite and its mineralogy was based on the mineralogical analysis of Murchison by \cite{bland2004} and \cite{howard2009}.}

However, there are the following differences compared to the meteorite: the organic component of volatile rich carbonaceous chondrites contains polycyclic aromatic hydrocarbons (PAHs), some of which are known carcinogens and mutagens. Our research suggests that rough chemical fidelity can be maintained by substituting much safer sub-bituminous coal as our organic analogue. A second safety consideration is the choice of serpentine group materials. While most serpentine polymorphs are quite safe, we avoid fibrous chrysotile because of its asbestos content. In the same vein, pyrite was substituted for troilite because of troilite's tendency toward combustion and/or explosion in small particle sizes \citep{yang2011}.

The simulant was produced by first sourcing the individual mineral components, which are then comminuted down to approximately 70~$\upmu$m in size. \ca{This size can be larger than the original CM meteorites with grain sizes ranging from sub-micron to micron \citep{britt2019}, and could potentially affect the thermal conductivity of the material.} The constituent minerals are then mixed together with water and sodium metasilicate, which acts as a binder. Approximately one ml of deionized water was added for every 4 grams of minerals. The resulting mixture is then cured at a high temperature ($\sim$80$^{\circ}$C) to dry and remove the water, then mechanically ground in a rock crusher to achieve the desired particle size distribution. \ca{The initial grain density of the simulant is about 2,750~kg~m$^{-3}$, which is in the range of the measured grain densities of the CMs, 2,570--2,870~kg~m$^{-3}$ \citep{britt2003}.} Next, we created regolith with simulated chondrite inclusions by adding glass spherules 600--800~$\upmu$m in size \ca{and 2,500~kg~m$^{-3}$ in density} so that the resulting mixture contained 15\% chondrules-analogues and 85\% minerals by volume, \ca{resulting to a final grain density of $\rho_\text{grain}$=2,710~kg~m$^{-3}$}. Mineral constituents, glass spherules, water, and sodium metasilicate were combined, as described above, and then packed into a rectangular cast with approximate dimensions of 95 $\times$ 95 $\times$ 44~mm \ca{in order to simulate the blocks present of the surface of Bennu and Ryugu}. The simulant was cast in and then cured at a high-temperature to remove moisture. Using a laboratory balance with precision $\pm$0.1~gr, we measured the mass of the target samples before the impacts, while their accurate dimensions (x,y,z) were measured using callipers. The ratio between mass and volume resulted in a density of the targets of $\rho_\text{bulk}$=1,980~kg~m$^{-3}$, very similar to the average of CMs \cite[2,200~kg~m$^{-3}$ from][]{2002BRITTast3}. \ca{The porosity, $\phi = 1 - \frac{\rho_\text{bulk}}{\rho_\text{grain}}$,  of the simulant blocks is calculated to be 26\%.}

We initially performed Thermogravimetry/Evolved Gas Analysis (TG/EGA) in order to measure the mass loss by volatile sublimation an its composition as a function of the temperature of the specimens. We found a that total volatile release of 11.2\% for the CM simulant, with the majority of the volatile content being the H$_{2}$O from the hydrated minerals. The mineralogy and bulk chemistry of our CM2 asteroid regolith simulant is reported in Tab.~\ref{table1}. The specimens were then shipped from Florida to France for further analysis. 

\begin{table*}
\centering
\caption{Mineralogy and bulk chemistry of CM Carbonaceous Chondrite Simulant (UCF/DSI-CM-2).}
\label{table1}
\begin{tabular}{l|l|l|l}
\hline
\hline
Mineral & Wt.\% & Oxide & Wt.\% \\
\hline
Mg-serpentine & 72.5  & SiO2 & 32.5\\
Magnetite & 10.4 & TiO2 & 0.3\\
Olivine & 7.8 & Al2O3 & 3.1\\
Sub-bituminous coal & 3.6 & Cr2O3 & 0.2\\
Pyrite & 2.6 & FeOT & 20.2\\
Pyroxene & 2.1 & MgO & 32.1\\
Siderite & 1.0 & CaO & 3.1\\
 && Na2O & 6.2\\
&&K2O & 0.2\\
&&P2O5 & 0.4\\
&&SO3 & 1.5\\
\hline
Total & 100.0 &Total & 100.0\\
\hline
\hline
\end{tabular}
\end{table*}

\section{Mechanical properties of simulant materials}
\label{test_mechanical}

\begin{figure}
\includegraphics[width=1.\columnwidth]{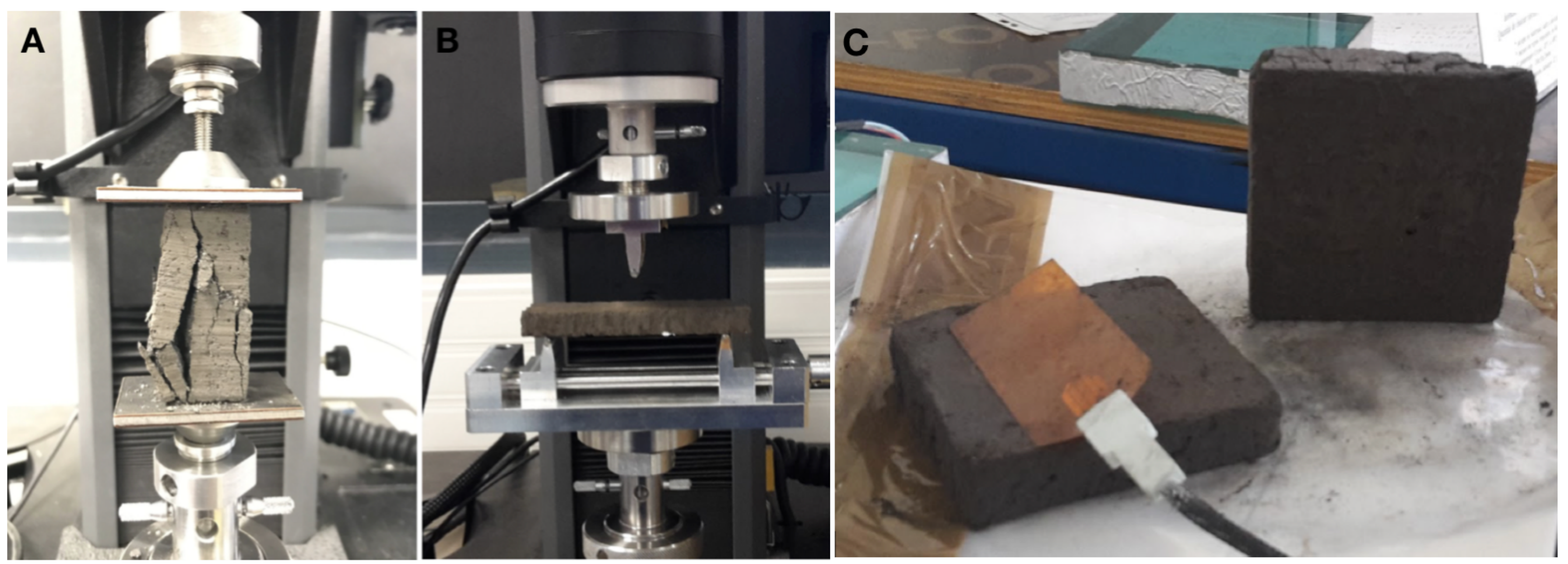}
\caption{Experimental devices for simple compression (A), 3-point bending test (B) and measurements of the thermal conductivity (C) .}
\label{measurements}
\end{figure}

Next, we performed experiments to measure the mechanical properties of the simulants.
Simple mechanical compression tests were carried out to determine: (1) the simple compression elastic modulus, $E_{c}$, and (2) the maximum compressive stress, $\sigma_{c}$ of our asteroid simulants. In addition, three-points bending tests were also carried out to determine (3) \ca{the flexural strength, $\sigma_{f}$, which can be related to the tensile strength (but see caveats below)}. In total, 12 flexural tests and 6 compression tests were performed \ca{(see Fig.~\ref{measurements})}. 

\ca{For the aforementioned mechanical tests,}\ the samples were prepared by cutting the simulants into parallelepipeds of roughly 20 $\times$ 20 $\times$ 50~mm for the compression tests and 10 $\times$ 25 $\times$ 75~mm for bending tests.
Parallelepipeds of 10 $\times$ 20 $\times$ 40~mm were also cut out of the main samples for additional bending tests. After the cutting, the samples were oven-dried at 378~K for at least 24 hours in dry air in order to remove the humidity from the Earth's atmosphere that have been absorbed by the simulants. We note that this heating process is too weak to drive off water present within the hydrated minerals as it has been shown that heating significantly above 373~K is necessary to alter the 2.7~$\upmu$m hydration spectral band of CM2 Murchison meteorite \citep{hiroi1996}. However this preparation improved the repeatability of the mechanical tests by limiting the influence of ambient humidity.

\ca{For brittle materials (such as the tested simulants), and for rocks in general, the compressive strength is significantly larger than tensile strength. Therefore, in a bending test, failure is triggered at the location of maximum tensile stress, namely at the vertical of the indenter on the opposite (lower) surface. At this location the stress is composed of a vertical component and a shear component. However, the shear stress is significantly lower than the tensile stress; this can be demonstrated analytically \citep{timoshenko1982}. For all the sample dimensions used in this study, we neglected the shear stress as it affects the failure stress only by a few percents in relative value, that is lower than experimental discrepancy between repeated measurements (see Tab.~\ref{table2}). Once we have neglected the shear stress, the point where the crack appears is in pure tension and therefore the flexural stress can be assimilated to the tensile stress. However, in a general way, the presence of defects in the material tends to cause the failure strength to be lower in a tensile test than in a bending test. The reason for this lies in the fact that in the bending test, only a small portion of the sample is loaded at the maximum tensile stress, whereas in the tensile test the stress is approximately homogeneous, so that the whole volume of the sample is submitted to the same stress. A larger volume implies more defects susceptible of triggering failure, thus reducing the tensile strength compared to flexural strength. As a conclusion, the flexural strength can be considered as an upper bound for the tensile strength.}

On the other hand, even for the samples with a higher aspect ratio, the \ca{flexural} modulus from the flexural tests was significantly different from that obtained from the simple compression ones, indicating that it is a combination of the compressive and \ca{flexural} moduli. Therefore, no conclusion could be drawn concerning the tensile elastic modulus.

\ca{In both bending and simple compression tests, the displacement $u$ of the press arm and the force $F$ applied where recorded together. The samples were loaded at 1~mm~min$^{-1}$. The compression stress was defined as $F/S$ ($S$ being the sample's cross-sectional area), the bending stress as $F\ell/4wt^2$, where $\ell$, $w$ and $t$ are the support span, width and thickness of the sample. The compression strain was defined as $u / L_0$, where $L_0$ is the initial sample length, and the bending strain as $6tu/\ell^2$.}

\begin{figure}
\includegraphics[width=1.\columnwidth]{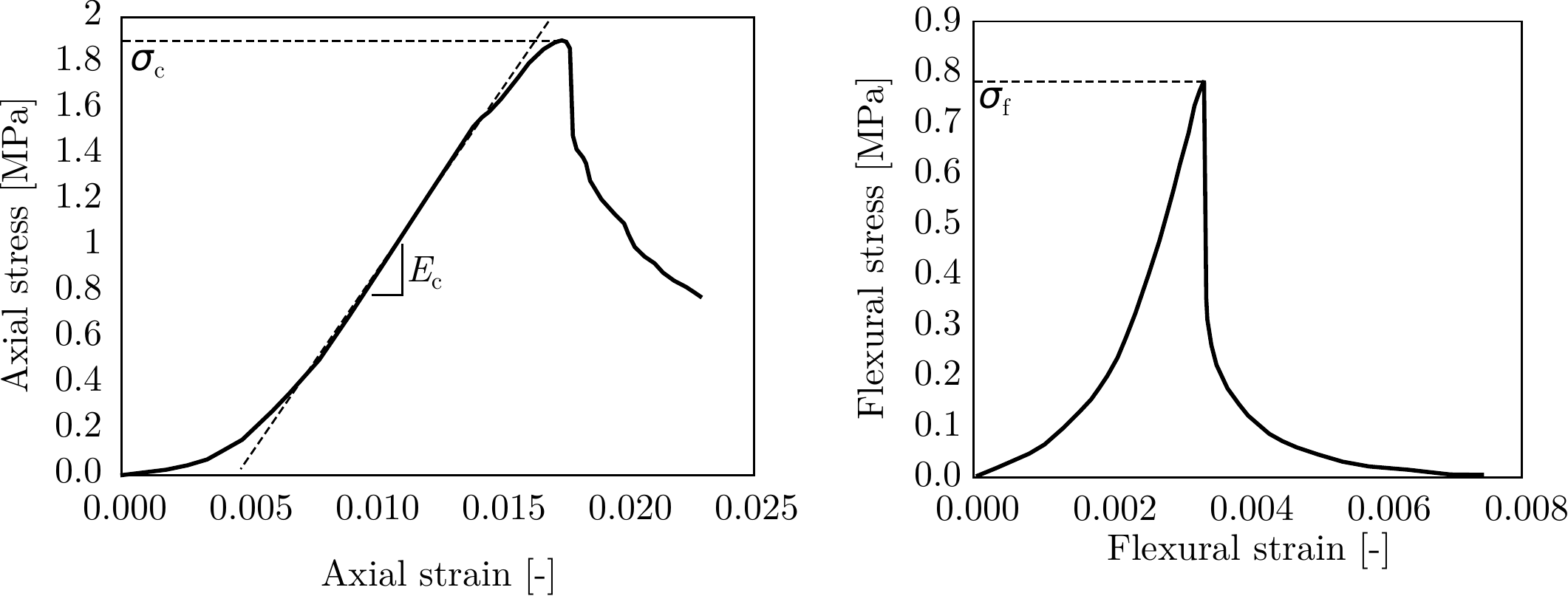}
\caption{Typical stress-strain curves with indication of compression strength $\sigma_{c}$ and elastic modulus $E_{c}$ (left) and the \ca{flexural} strength $\sigma_{f}$ (right).}
\label{tests}
\end{figure}

In Fig.~\ref{tests} we present the average stress-strain curves for flexural and compression tests. The maximum of the curves give the compressive $\sigma_{c}$ and flexural $\sigma_{f}$ strengths of the simulant. As described before, we approximate the tensile strength of the material to be equal to the flexural strength for this particular experiments. For the compressive stress, $E_{c}$ is the slope of the linear part of the stress-strain curve.
Samples were cut in two perpendicular directions to assess the anisotropy of the \ca{simulant} material. However, no significant differences were observed in the resulting measurements of the values for $\sigma_{f}$, $\sigma_{c}$, and $E_{c}$. Therefore, no further mention of anisotropy will be made, and we processed all the results of the mechanical tests assuming the samples to be equivalent with respect to anisotropy of the mechanical properties.

Due to the weak and brittle nature of the \ca{simulant} material, it was difficult to obtain perfectly parallel surfaces for the samples. Broken corners, holes, bad surface roughness and poor parallelism between the sample's faces affect the repeatability of the test. This inaccuracy, to which should be added the inhomogeneities of the material itself, induced an uncertainty in the measurement. Therefore, it was assumed that the measured quantities $\sigma_{f}$, $\sigma_{c}$ and $E_{c}$ followed a normal distribution and accordingly, a 90\% confidence interval was calculated. The results of the mechanical tests are summarised in Tab.~\ref{table2}. 
The smaller samples (10 $\times$ 20 $\times$ 40~mm), were used for the bending tests in order to study if the results are significantly affected by aspect ratio of the samples. We find that results are consistent regardless the dimensions of the samples. 

\begin{table*}
\centering
\caption{Mechanical properties of CM Carbonaceous Chondrite Simulant (UCF/DSI-CM-2).}
\label{table2}
\begin{tabular}{l|lll}
\hline
\hline
& $E_{c}$ (MPa) &  $\sigma_{c}$ (MPa) &  $\sigma_{f}$ (MPa) \\
\hline
mean value & 151.67 & 1.82 & 0.72\\
\hline
90\% conf. interval & 17.48 & 0.17 & 0.07\\
\hline
\hline
\end{tabular}
\label{tab:mech_results}
\end{table*}

\section{Thermal properties of simulant materials}
\label{test_thermal}

We also measured the thermal conductivity of our asteroid \ca{simulants} in the laboratory. We adopted the hot wire method, using a probe developed by the \emph{Centre Scientifique et Technique du Batiment} CSTB (Scientific and Technical Center for Building), in Grenoble, that has a typical accuracy of 5\%. To do this, one 95 $\times$ 95 $\times$ 44~mm sample was cut in two pieces of about 95 $\times$ 95 $\times$ 20~mm, to place the probe between them, as shown in Fig.~\ref{measurements}C. The measurement was repeated three times and attention was paid to limit the maximum temperature transmitted to the sample to be below 310~K. The three measurements were done at ambient temperature (the average temperature was 295~K) and atmospheric pressure in air. The measurement was quite repeatable, showing a value of thermal conductivity being between 0.43 and 0.47~W~m$^{-1}$~K$^{-1}$.

\section{\ca{Impact experiments: ejecta, crater sizes and morphology}}
\label{experiments}

In order to study the response of \ca{our asteroid simulant} material to collisions with particles at typical impact speeds occurring in the asteroid Main Belt, we performed a series of laboratory hypervelocity impact experiments. We used the facilities of the Impact Lab of the University of Kent in Canterbury (UK). The main instrument used here is a 2-stage light-gas gun (LGG), which can achieve speeds up to 7.5~km~s$^{-1}$, while the typical impact speed in the Main Belt is around 5~km~s$^{-1}$ \citep{bottke1994}. In our experiments we used spherical projectiles made of stainless steel, with density of 7,870~kg~m$^{-3}$. 
\ca{It should be noted that this value is much higher than the typical densities of the micrometeoroids that impact asteroid boulder in the Main Belt and near-Earth space, which are of cometary and asteroidal origin \citep[][and references therein]{nesvorny2010}. However, by choosing a high density projectile we wanted to ensure the production of a substantial amount of ejecta, even at the lower speeds.}

\begin{figure}
\includegraphics[width=1.\columnwidth]{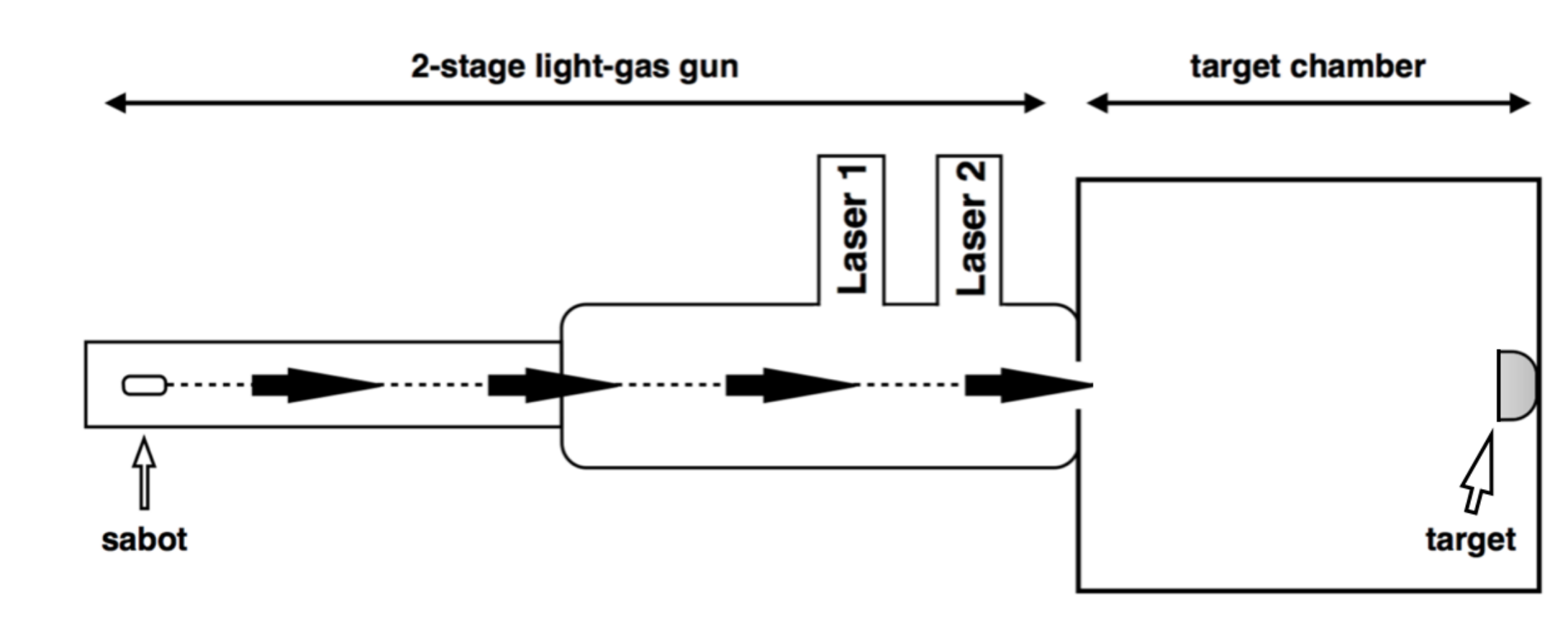}
\caption{Schematic of the 2-stage light-gas gun of the University of Kent.}
\label{gun}
\end{figure}

The configuration of the shotgun cartridges used in the first stage, and gas pressure in the second stage of the LGG, were calibrated to produce controlled impact speeds varying between 0.8 and 5~km~s$^{-1}$. The actual impact speed for each shot is measured by timing the passage of the projectile through two laser curtains placed down-range in the gun and is indicated in Tab.~\ref{table3}.
For all the tested impact speeds we used projectiles with diameter 1~mm with mass of 0.0041~gr. The shot with impact velocity of 4.7~km~s$^{-1}$ was repeated tree times using stainless steel projectiles with diameters of 1, 1.5 and 2~mm, the latter two having masses 0.0139~gr and 0.0330~gr, respectively. This procedure was carried out in order to compare impacts at similar speeds, but with different kinetic energies. The uncertainty of the measurements of projectile diameters is about 10~$\upmu$m.

Targets comprised of the blocks of CM simulants prepared as described in section 2 with approximate dimensions 95 $\times$ 95 $\times$ 44~mm thickness.
The target samples were placed in a holder which was mounted on the door of the target chamber and were impacted horizontally (Fig.~\ref{gun}). The impact chamber was evacuated prior to each shot, the air pressure inside the target chamber was measured and is reported in Tab.~\ref{table3}. 
At the bottom of the target chamber were placed several layers of aluminium foil that enabled the collection of the ejected fragments after each shot for later inspection. 

\begin{figure}
\includegraphics[width=1.\columnwidth]{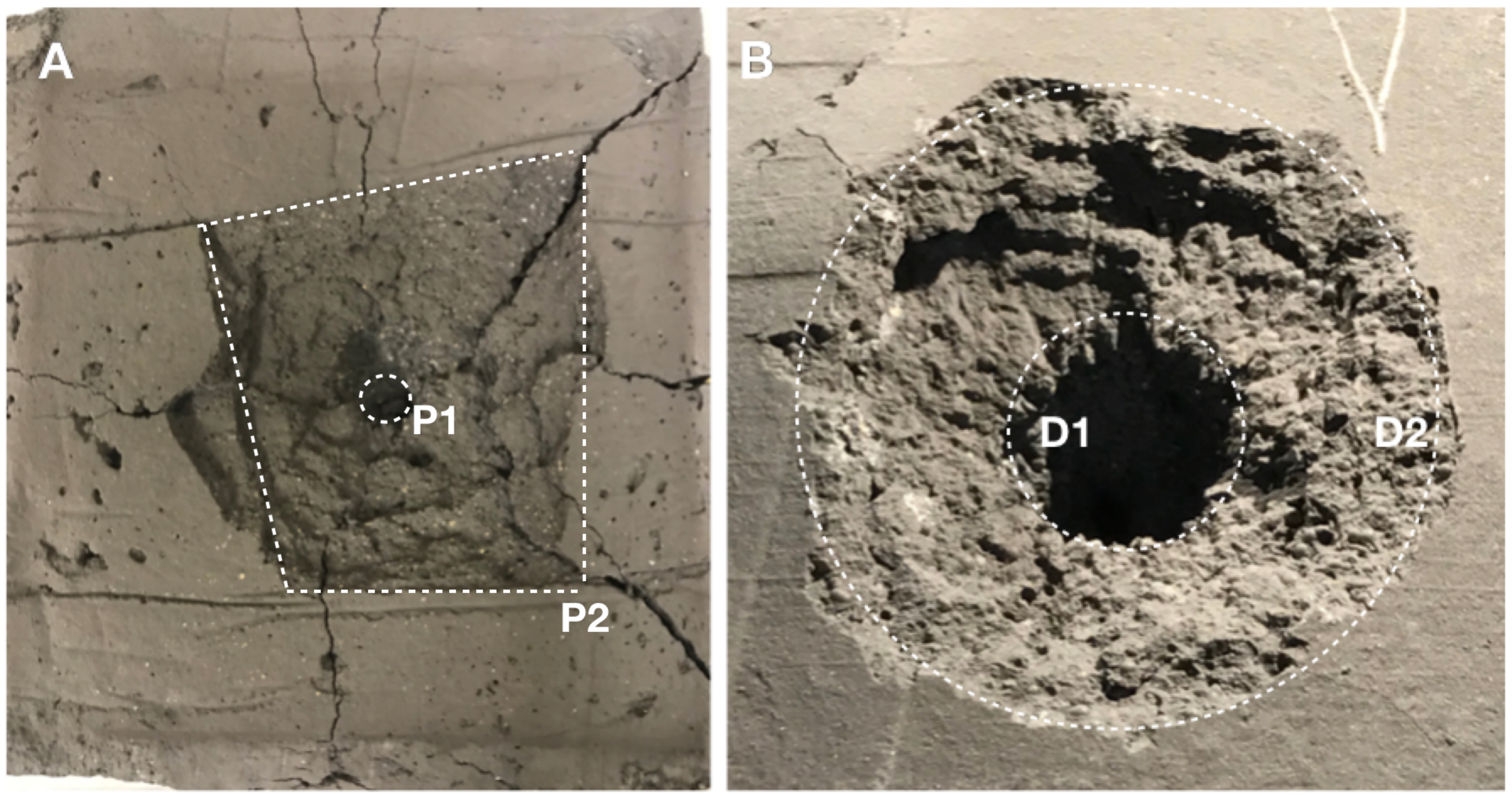}
\caption{The produced craters of two experiments. (A) Crater of the shot G181109\#3 with impact speed 4.71~km~s$^{-1}$ and 2~mm size projectile. The area between P1 and P2 was used to estimate the roughness of the cratered area; (B) Crater of the shot G181109\#2 with impact speed 4.81~km~s$^{-1}$ and 1~mm size projectile. Both images show the central crater (inner white dashed circle, in B panel with diameter D1), and the outer spallation area (zone zone in between D1 and D2) with incipient spallation. Diameters, D1 and D2, and depths are in Tab.\ref{table3}.}
\label{craters}
\end{figure}

With these experiments we primarily aimed to examine the crater sizes, shapes and morphology, along with the type of the produced ejecta (e.g. monominerallic, multiminerallic). After each shot the targets were removed with great care from the chamber door in order to preserve the shape of the produced crater. For each crater it was measured the total depth $d$, the crater diameter $D1$ and the larger diameter encompassing the area of the spall $D2$, which are presented in Tab.~\ref{table3}. The spallation appears when there is tensile failure as the compressive stress is reflected at the surface and becomes tensile stress. In addition, we present the $d/D1$ and $d/D2$ ratios of each crater. In the case of three impacts where the impact speed is constant but the impact energy is different the $d/D1$ and $d/D2$ ratios appears also constant.
In the case of the shot G181109\#3 the surface of the target was not smooth but had large cracks that were formed during the production phase. The resulted crater was not circular, but had an irregular shape following those cracks (see Fig.~\ref{craters}A). 

\begin{figure}
\includegraphics[width=1.\columnwidth]{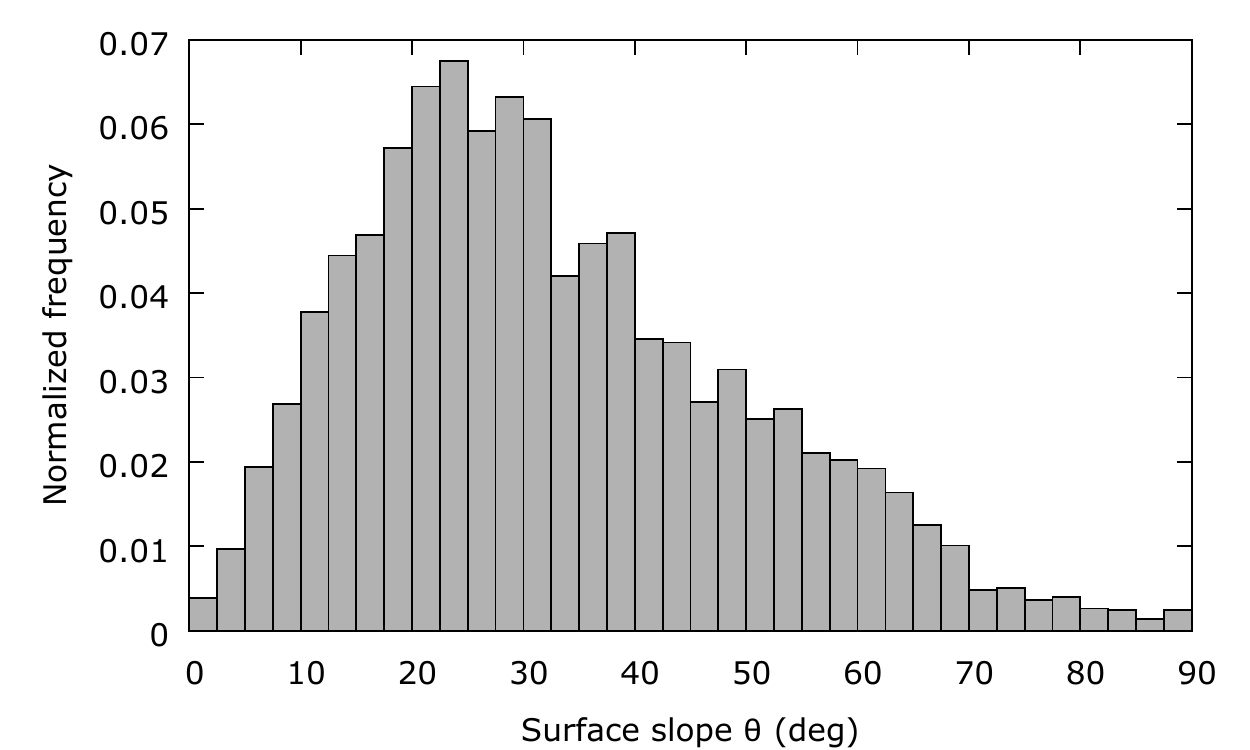}
\caption{\ca{The distribution of the surface slopes within the damaged area of the surface of the simulant after the shot G181109\#3 (see Fig.~\ref{craters}A) as measured from the 3D scan of the simulant.}}
\label{F:a_theta}
\end{figure}

\ca{Next we estimated the roughness of the impacted surface. For this measurement we used the target with the largest formed crater (with a diameter of $\sim$5~cm) from the shot G181109\#3 (Fig.~\ref{craters}A). The cratered simulant block was scanned using a commercial 3D software (Qlone) and the object exported into a \emph{wavefront (.obj)} file format representing a mesh of 344,312 triangular facets (with a side-length of 0.3~mm). A region of interest with 32,911 facets corresponding to the surface of the simulant damaged by the impact and without the central hole (region between polygons P1 and P2 of Fig.~\ref{craters}A) was selected to perform the roughness analysis. The mean plane of the surface was calculated along with the angle between the normal of this plane and the i-th facet that we call $\theta_i$. Following \cite{hapke1984}, we characterise the roughness of the cratered region of the simulant by its mean surface slope $\bar{\theta}$, which is defined as: $\tan(\bar\theta) = (2/\pi) \int^{\pi/2}_{0} \tan(\theta) a(\theta) d\theta$ where $a(\theta)$ is the distribution of the surface slope normalised such that $\int_0^{\pi/2} a(\theta) d\theta = 1$ (as displayed in Fig.~\ref{F:a_theta}). We find that $\bar\theta$ = 32$^\circ$ for the surface of our sample damage by the impact. In section 6, we compare this value with the Ryugu boulder's surface from \cite{grott2019} data.}

\begin{figure}
\includegraphics[width=1.\columnwidth]{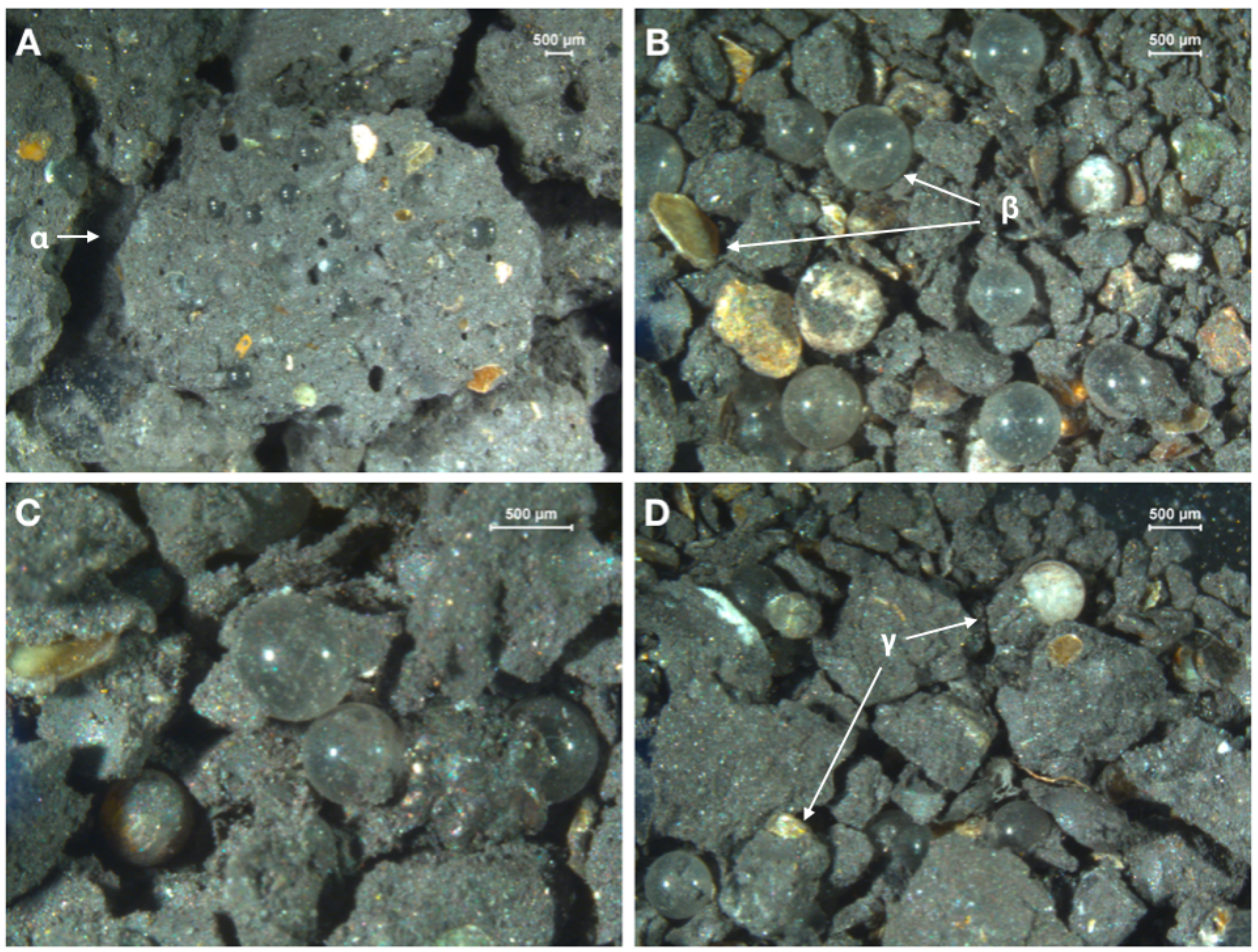}
\caption{Parts of collected ejecta produced at four impact experiments. In each case we found multimineralic large fragments ($\alpha$), monomineralic fragments ($\beta$) and individual inclusions that were attached to a small amount of CM matrix.}
\label{ejecta}
\end{figure}

For four out of six impact experiments ejecta was recovered. Ejecta could not be collected for the low speed shots S181106\#1 and G181105\#1 as the craters were very small and the excavated crater mass was comminuted to a fine powder, which was impossible to collect using the setup. For the rest of the shots, G181105\#2, G181109\#1, G181109\#3, G181109\#2, that were performed at higher and narrow speed range between 4.07 and 4.81~km~s$^{-1}$, we see similar characteristics of the ejecta. First of all ejecta has a size range from centimetres to sub-microns. Furthermore, as can be seen from the different panels of Fig.~\ref{ejecta}, the ejecta consist of both multimineralic and monomineralic fragments, where the glass inclusions and the pyrite dominate the monomineralic population of sub-mm-sized fragments.

At any impact speed, the largest fragments, such as the one at the centre of Fig.~\ref{craters}A, which measures 8~mm along its longest axis, are multiminerallic: the glass spherules and the pyrite are still embedded in the CM-like matrix. The multiminerallic nature of impact fragments was expected as described in the introduction of this work. Additionally, we see many small fragments which consist primarily of a glass spherule or a pyrite attached to a smaller piece of the CM matrix.
On the other hand, several smaller fragments ($\sim$500~$\upmu$m in size) are clearly monominerallic (see morphological units $\alpha$, $\beta$ in the Fig.~\ref{ejecta}).

\begin{figure}
\includegraphics[width=1.\columnwidth]{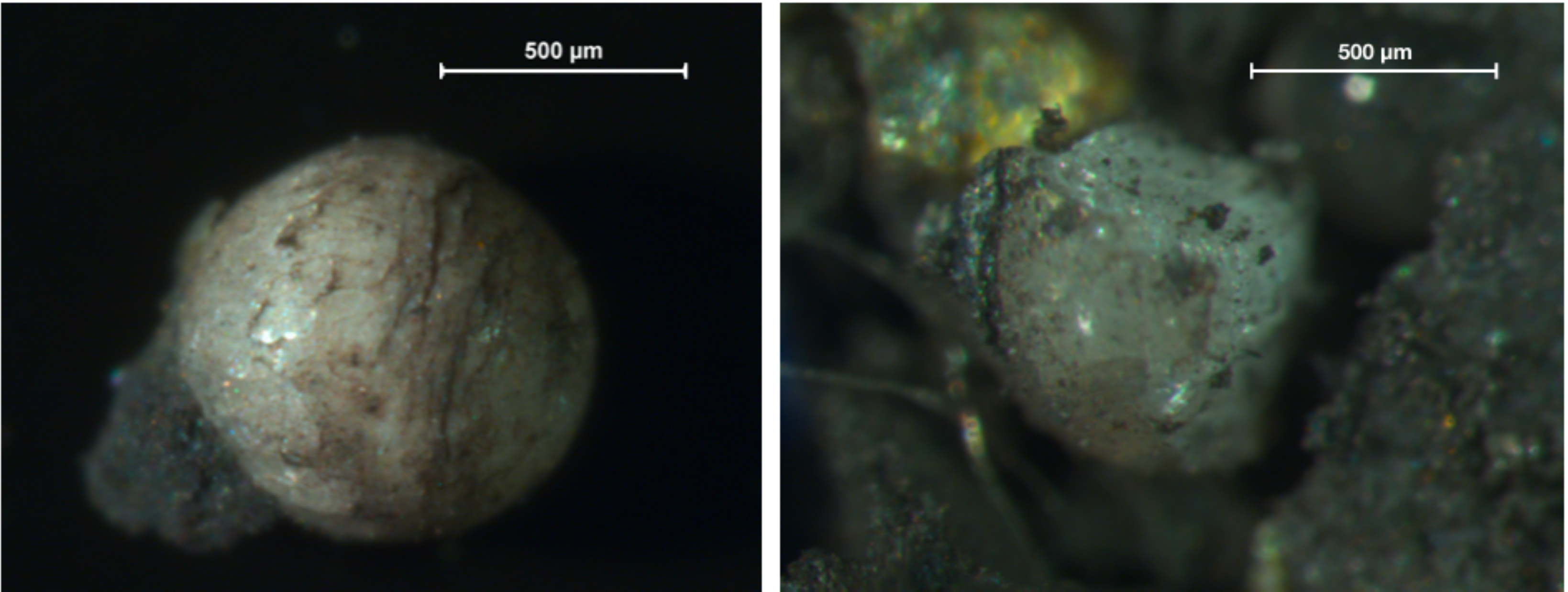}
\caption{Examples of a shocked, shuttered and broken spheres that were used as inclusions in the CM matrix.}
\label{sphere}
\end{figure}

In all the ejecta material collected we discovered severely shocked individual glass spherules (Fig.~\ref{ejecta}). These spherules have white colour indicating that are totally shattered and very possibly originate from the centre (or close to the centre) of the impact (see an example in Fig.~\ref{sphere}). 

\begin{table*}
\centering
\caption{Summary of the impact experiments where are given: the speed of each impact ($V_{imp}$), the air pressure inside the impact chamber ($P$), the diameter of the projectile ($D_{proj}$), the impact energy ($E_{imp}$), the depth ($d$) and diameters of pit and spall ($D1$ and $D2$) of each crater. In addition, are given the depth-to-diameter ratio for both cases ($d/D1$ and $d/D2$) and the spall-to-pit ratio ($D2/D1$).}
\label{table3}
\begin{tabular}{l|llllllllll}
\hline
\hline
Shot ID & $V_{imp}$ &  P & $D_{proj}$ &  $E_{imp}$ &  $d$ & $D1$ & $D2$ & $d/D1$ & $d/D2$ & $D2/D1$  \\
	& (km~s$^{-1}$) & (mbar) & (mm)	& 	(kJ)	&	(mm)		&	(mm)		&	(mm)		& &&\\
\hline
S181106\#1	& 0.85&0.76 	& 1.0 & 1.48	& 7.5 & 1.97 & 4.20 & 3.80 & 1.78 & 2.13\\
G181105\#1	& 2.13&0.90	& 1.0 & 9.25	& 9.0 & 2.95 & 12.05 & 3.05 & 0.74 & 4.08 \\
G181105\#2	& 4.07&0.70	& 1.0 & 33.79	& 4.0 & 6.78 & 23.25 & 0.59 & 0.17 & 3.43\\
G181109\#1	& 4.70&0.49	& 1.5 & 152.13	& 12.0 & 9.24 & 27.20 & 1.29 & 0.44 & 2.94\\
G181109\#3	& 4.71&0.25	& 2.0 & 362.16	& 17.0 & 14.00 & 51.62 & 1.21 & 0.33 & 3.7\\
G181109\#2	& 4.81&0.55	& 1.0 & 47.00	& 5.5 & 5.92 & 18.88 & 0.93 & 0.29 & 3.2\\
\hline
\hline
\end{tabular}
\end{table*}

\section{Discussion}

The first action we did was to measure the mechanical properties of the \ca{simulant} material and compare them to the properties of some known weak carbonaceous meteorites. In particular, the compressive strength of the \ca{simulant} is found to be 1.82 $\pm$ 0.17~MPa, at least an order of magnitude smaller than the one measured for the CM2 meteorites Murchison \citep{miura2008} and Sutter's Mill \citep{jenniskens2012}. The value of the compressive strength of our simulant is closer to the one of the C2 ungrouped Tagish Lake meteorite that was found to be around 0.7~MPa.
The \ca{flexural} strength of the \ca{simulant} was measured to be 0.72 $\pm$ 0.07~MPa: \ca{this value, according to our approach presented in section 3, represents an upper limit for the tensile strength of the simulant and thus probably even lower than the tensile strength of the CI meteorite Ivuna (0.7 $\pm$ 0.2~MPa) and C2 Tagish Lake (0.8 $\pm$ 0.3~MPa) as given by \cite{tsuchiyama2009}}. However, the \ca{inferred tensile strength of our simulant is at least one magnitude lower the one derived for CM2 meteorite Murchison and Sutter's Mill, and Murray \citep[8.8 $\pm$ 4.8~MPa;][]{tsuchiyama2009}.} Here we have to mention that our experiments were conducted in air, but this does not affect the measurements since the compressibility of the air is below that of the solid materials.

\begin{figure}
\includegraphics[width=1.\columnwidth]{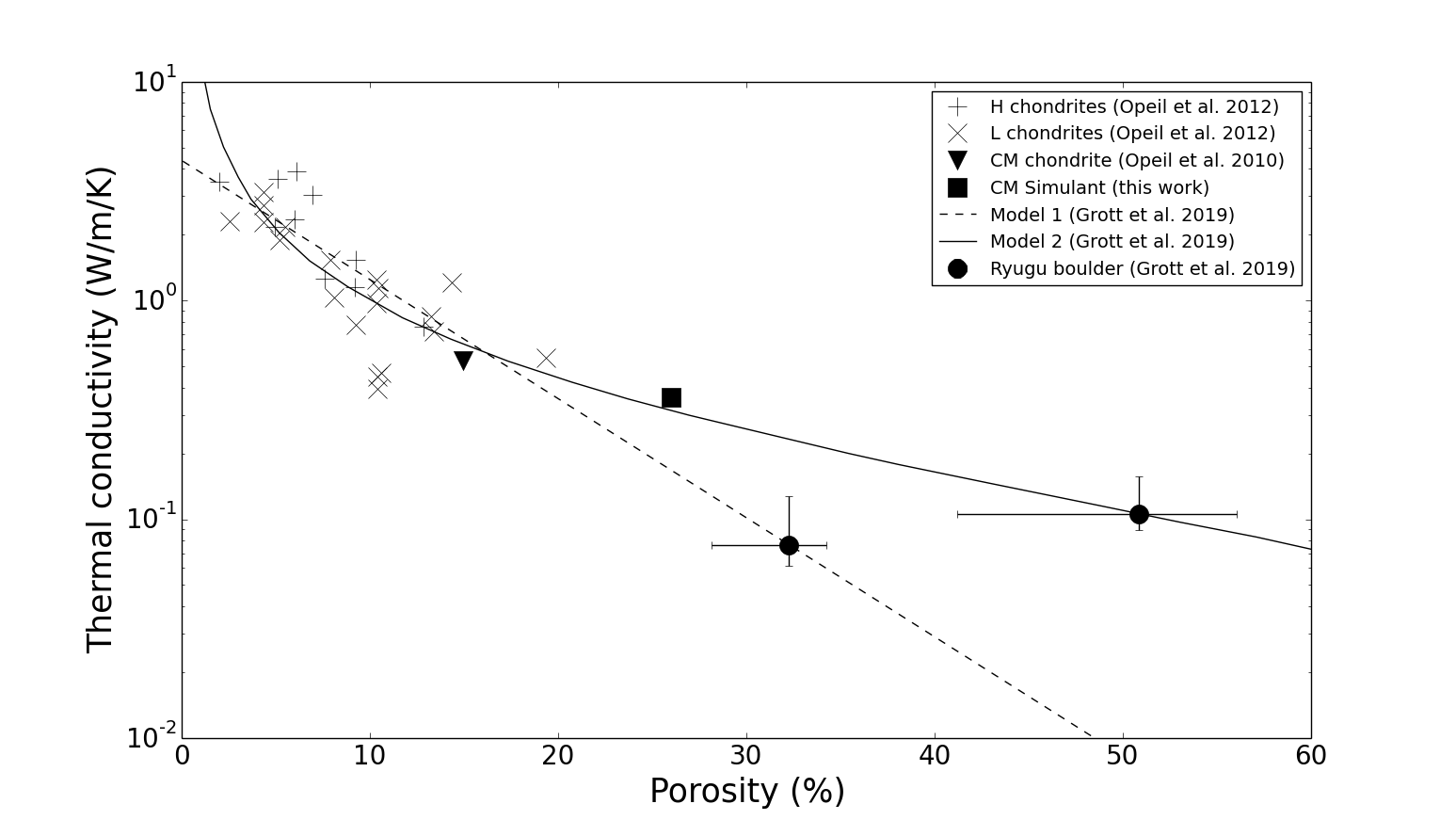}
\caption{Thermal conductivity of H, L and CM meteorites, in comparison with the measured thermal conductivity of the surface boulders of Ryugu and with our asteroid surface simulant. Adapted from \cite{grott2019}.}
\label{kappa}
\end{figure}

\ca{The measured thermal conductivity $\kappa$ of the CM carbonaceous chondrite Cold Bokkeveld at the temperature of 200~K is 0.50~W~m$^{-1}$~K$^{-1}$ \citep{opeil2010} and is temperature dependent ($\kappa$-values were measured with the meteorite samples inside a cryostatic chamber in the temperature range between 300~K and 5~K and at pressure $<1.33 \times 10^{-4}$~Pa). On the other hand, the $\kappa$-value of our simulant was measured to be between 0.43 and 0.47~W~m$^{-1}$~K$^{-1}$ at room temperature (295~K) in air. However, because thermal conductivity is temperature dependent, we need to correct the measured $\kappa$-values for Cold Bokkeveld and our simulant to the same temperature at which Ryugu's boulder thermal conductivity was derived. The boulder of Ryugu was calculated to have a $\kappa$=0.06--0.16~W~m$^{-1}$~K$^{-1}$ at 230~K \citep{grott2019}. Using the temperature dependence curve of the Cold Bokkeveld CM meteorite, we make the correction on the basis of a linear fit to that data of Fig.~3 of \cite{opeil2010} for which temperature is greater or equal to 150~K and we get $\kappa$=0.53~W~m$^{-1}$~K$^{-1}$ at a temperature of 230~K for the CM Cold Bokkeveld. Applying the same temperature dependence of the Cold Bokkeveld to our thermal conductivity measurement after having subtracted the air thermal conductivity (0.03~W~m$^{-1}$~K$^{-1}$), we result in $\kappa$=0.36~W~m$^{-1}$~K$^{-1}$ at 230~K. This shows that the $\kappa$ of our simulant is lower than the one measured for Cold Bokkeveld. The bulk porosity $\phi$ of our simulant is higher than those of meteorites, placing it in between the value of Cold Bokkeveld and the one estimated for the boulder of Ryugu. The thermal conductivity vs. porosity data of our simulant, the Cold Bokkeveld meteorite and Ryugu's boulder plot along the prediction of the Model 2 ($\kappa$ = 0.11 $\times$ (1-$\phi$)/($\phi$)) of \cite{grott2019} and are presented in Fig.~\ref{kappa}.} 

\ca{We therefore summarise that the mineralogy, water content and amount of inclusions of our simulant is close to CM meteorites, while the thermal conductivity and mechanical properties are in between those of CM meteorites and the boulder of Ryugu. In the light of the new space mission data, it appears that we miss asteroid material from our meteoritic collection, perhaps not in terms of composition, but very likely in terms of thermo-mechanical properties. This is very reasonable as Earth's atmosphere filters out the weakest fragments. Therefore, the novelty behind this work is not to precisely reproduce the existing sample of weak carbonaceous meteorites but to mimic more accurately the surfaces of carbonaceous asteroids.}

The morphology of the craters \ca{created by hypervelocity impact of mm-sized projectiles on our simulants is typical of those} of brittle materials \citep{dufresne2013}. All craters show a central \ca{pit} with diameter $D1$ and a spallation area around. There was a high density difference between \ca{our simulant} material (1,980~Kg~m$^{3}$) and the steel projectiles (7,870~Kg~m$^{-3}$), which resulted into large spalling areas and also large depth-to-diameter ratio. Our impact craters simulate the result from micrometeorite bombardment and the spall-to-pit diameter ratio ($D2/D1$) ranges between 2 and 4. \ca{The latter is slightly larger than the one for} lunar microcraters that have spall-to-pit ratio around 3 and \ca{more precisely falling} in the range 1.5--3, as it is reported by \citep{nakamura2017}. The depth of our craters is large (see Tab.~\ref{table3}) and this can be explained by the large ratio of projectile-to-target density. In addition, our depth-to-diameter ratio is also large, ranging between 0.2 and 0.7. 

On the other hand, the depth-to-diameter ratio derived from the large craters (with diameters between $\sim$10 and 290~m) found on asteroid Ryugu was measured to be 0.14--0.2 \citep{sugita2019}, and is smaller than the ones we derive in this work. \ca{This difference in crater depths between our experiments and Ryugu's craters can be explained by a series of observations: \cite{sugita2019} explain the shallow craters on Ryugu as due to the presence of loosely unconsolidated material that could have moved, producing landslides, and is also responsible for the observed crater floor morphologies. Moreover, another reason to reduce the crater sizes is the presence of m-sized boulders, comparable to the impactors' sizes, which absorb a large portion of impact energy. As a conclusion, our experiments are more comparable with the meteoroid impacts on the surface boulders of Ryugu or Bennu which form mini-craters (also known as cavities; R. Ballouz, private communication). These mini-craters have been detected in high spatial resolution images from the detailed survey phase of the OSIRIS-REx mission on boulders of Bennu. The cavities have diameters ranging from about 3 to 30~cm \citep{ballouz2019}. A quantitative comparison between the properties of our experimental craters and the mini-craters will be possible when spacecraft data will be publicly available. 
Moreover, our experimental craters resemble, at least morphologically, the surface at similar cm-scales that MASCOT revealed on Ryugu boulders \citep{grott2019,jaumann2019}. The roughness at a scale smaller than the MARA footprint -- about 5~cm in diameter -- of the Ryugu boulder observed by MASCOT can be estimated from the best fit parameters of the thermal model used by \cite{grott2019} to interpret MARA radiometric measurements. In particular, \cite{grott2019} find that a thermal model with 0.34 of the surface area observed by MARA covered by hemispherical craters produce the best fit to the data. From the crater opening angle ($\gamma_C$) and the crater areal density ($\rho_C$) it is possible to calculate the mean slope of surface $\bar\theta$ as described by \cite{Delbo2004PhD} and shown by \cite{hanus2018}. We find that $\gamma_C$=90$^\circ$ and 
$\rho_C$=0.34 yield $\bar\theta$=28.7$^\circ$. This value is very similar to the mean surface slope of the region of our simulant damaged by the hyper-velocity impact that we performed in the laboratory ($\bar\theta$=32$^\circ$). We thus conclude that the degree of roughness of the surface created by the impact on our simulant and that observed by MASCOT/MARA at very similar spatial scale are very similar. It is thus possible that MASCOT observed a surface resulting from the bombardment of micrometeorites over the age of Ryugu.}

The large multimineralic fragments, like the one in Fig.~\ref{ejecta}A, most probably originate from the spalling area, where the shock was already disseminated. On the other hand the shuttered and broken glass inclusions should come from the central crater. \ca{This idea could be supported by recent hypervelocity impact experiments conducted on L ordinary chondrite meteorites \citep{michikami2019} where the impact induced cracks appear to grow through the chondrules that are located near the impact point, and around the chondrules that are located away from the impact point. Although there are indications for a difference in the crack propagation, the authors did not measure a strong correlation between the type of crack propagation in chondrules and the distance of those chondrules from the impact point concluding that for the near-Earth asteroid Itokawa impacts is the main mechanism of regolith production with preference to multimineralic fragments. On the other hand, early impact experiments on basalt with olivine inclusions showed a preference in the generation of olivine dominated fragments, indicating a crack propagation around the density boundaries between olivine and basalt matrix \citep{durda1999}.
From our impact experiments, where both monomineralic and multimineralic fragments are present, does not lead to an obvious separation of the production mechanism and thermal cracking is not the only process that produces monomineralic fragments. By broadly combining these three different impact experiments we could argue that there should be a difference on how strongly are bound the inclusions to the matrix of different materials, natural and simulant.}

The key characteristic for a clear answer, when sample will return from Ryugu and Bennu, could be the presence of shocked fragments where here are represented by the shuttered glass inclusions.

\section{Conclusions}
We created in the laboratory a mineralogically analogue material to the CM carbonaceous meteorites but with mechanical and thermal properties intermediate between those of the CM meteorites, that we receive on Earth, and those measured by space missions for the boulders on the C-complex near-Earth asteroids Ryugu and Bennu. We added to the matrix of a CM-like composition spherical glass beads in order to simulate the multiminerallic nature of known CMs due to the presence of chondrules and pyrite, the latter being occasionally present in CM meteorites, such as in the case of the Murchison meteorite \citep{fuchs1973}.

We measured the compressive and tensile (flexural) strength of the \ca{simulants}, that we found to be around 1.8 and 0.7~MPa, respectively (with a 0.17 and 0.07~MPa 90\% confidence interval).
We also measured the thermal conductivity (in air) of the \ca{simulants} at room temperature (295 K), which resulted in a value between 0.43 and 0.47 W~m$^{-1}$~K$^{-1}$. The values of the mechanical and thermal properties are intermediate between the typical ones of carbonaceous chondrites \citep{brown2002, miura2008, opeil2010, jenniskens2012} and those measured and/or inferred for the boulders of the asteroid Ryugu \citep{grott2019,sugita2019}. Preliminary observations of the OSIRIS-REx mission indicate that also the boulders covering the surface of Bennu have similarly low thermal conductivities and likely low mechanical strengths \citep{2019NatAs...3..341D}.

We performed a series of hypervelocity (0.8--5~km~s$^{-1}$) impact experiments in order to study how these materials fragment, simulating their response to collisions of micro-meteorites in the asteroid Main Belt. We determined that craters resulting from these impacts have typical morphologies, with large depth-to-diameter ratios that is characteristic of impacts on porous media. Morphologically, the surface resulting from the impact on the \ca{simulants} present a degree of roughness that qualitatively \ca{and quantitatively} resembles those observed by MASCOT on the boulders of Ryugu \citep{grott2019}.

We find that the material fragmented and ejected by an impact at a velocity of 4--5~km~s$^{-1}$ contains both multi- and monominerallic fragments, including single glass spherules that were easily separated from the matrix by the impact shock (we had originally embedded these glass spherules in the matrix of our \ca{simulants} to mimic chondrules). 

The results from this study indicate that it could be more difficult than predicted (see introduction) to distinguish grains produced from thermal fragmentation from those created by the impact of micro-meteorites in the regoliths of asteroids that have CM-like composition with mechanically weak surface boulders. This is because it was originally predicted that thermal fragmentation should produce monominerallic fragments \cite{hazeli2018} than impacts would do. 
A further work will investigate the impact effects on CI-like materials and also the effect of thermal cracking on both CM- and CI-like simulants with weak mechanical strengths similar tho those inferred for boulders on Ryugu. 

\section*{Acknowledgements}

This work was supported by the BONUS QUALIT\'E RECHERCHE Lagrange (BQR) 2017 and partially by the ANR ORIGINS (ANR-18-CE31-0014). The work of C.A. was supported by the French National Research Agency under the project ``Investissements d'Avenir" UCA$^\text{JEDI}$ with the reference number ANR-15-IDEX-01. This work was supported by the Programme National de Plan\'etologie (PNP) of CNRS/INSU, co-funded by CNES and by the program ``Flash!" supported by Cr\'edits Scientifiques Inditatifs (CSI) of the Universit\'e Nice Sophia Antipolis. D.B. and C.S. were supported by the NASA Solar System Exploration Research Virtual Institute cooperative agreement number NNA14AB05A to the Center for Lunar and Asteroid Surface Science.Authors would like to thank C. Pompeo from the CSTB (Grenoble) for his help with the measurement of the thermal properties. The laboratory LEGI is part of the LabEx Tec 21, supported by the project ``Investissements d'Avenir" with the reference number ANR-11-LABX-0030.

\bibliographystyle{model5-names}\biboptions{authoryear}
\bibliography{references,mypapers}

\end{document}